# Principal component analysis for v = 5 / 2 fractional quantum Hall states

Zhe Zhang[a], Jiajie Qiao[a], Xiaoliang Wu[a], Shaowen Yu[a], Qin Jin[*,a]

[a]College of Physics Science and Technology, Yangzhou University, Jiangsu 225009, China

[*]Corresponding Authors: jinqin@yzu.edu.cn

**Abstract**

For the special single-layer fractional quantum Hall system with a filling factor of 5/2, which has an even denominator, this paper uses principal component analysis (PCA) to study its behavior under the breaking of particle-hole symmetry. By introducing a model three-body potential to represent the mechanism of particle-hole symmetry breaking, the paper finds that the 5/2 system evolves into two types of special topological quantum states with non-Abelian statistics as the strength and direction of the three-body potential vary. The transition points of these states correspond to the particle-hole symmetric pure Coulomb interaction system. Our results validate the applicability of machine learning as a new research tool in fractional quantum Hall systems. Furthermore, machine learning directly analyzes the raw wave functions, without relying on prior empirical theoretical assumptions and models, making it applicable to a broader range of fractional quantum Hall systems experiencing phase transitions due to particle-hole symmetry breaking.

## 1. Introduction

The fractional quantum Hall (FQH) effect has emerged as a fundamental subject in condensed matter physics, as it reveals novel quantum phases that lack classical counterparts. Among the various FQH states, the 5/2 filling factor stands out due to its potential to host exotic topological quantum states with non-Abelian statistics. These non-Abelian anyons are of significant interest for their potential application in fault-tolerant quantum computing, where quantum information can be stored in a way that resists local perturbations. The possibility of using such topological states for designing robust qubits has made the study of the 5/2 state a central focus in the quest for quantum computing technologies. Theoretical efforts to explain the nature of the 5/2 FQH state have centered on two competing candidate wave functions: the Pfaffian (Pf) state and the anti-Pfaffian (APf) state. Both of these states are believed to exhibit non-Abelian statistics, offering considerable value for topological quantum computing applications. However, these states are not symmetric under particle-hole transformations, whereas the system's Hamiltonian, in the absence of Landau level mixing, is symmetric. This discrepancy presents a significant theoretical challenge: although the Hamiltonian preserves particle-hole symmetry, the observed ground states, likely corresponding to either the Pf or APf states, do not. Therefore, clarifying the mechanisms that break particle-hole symmetry in the 5/2 system is crucial for understanding the emergence of these asymmetric topological states.

One such mechanism for breaking particle-hole symmetry is Landau level mixing, which has been shown to play a crucial role in driving the ground state of the system towards either the Pf or APf state. In addition, previous numerical studies have demonstrated that introducing a three-body interaction potential is an effective approach for breaking this symmetry. The three-body potential enables the system to favor either the Pf or APf state depending on the direction and strength of the potential. However, these studies have primarily relied on model wave functions that are specifically tailored to either the Pf or APf state. This reliance on specific models limits the ability to explore the full range of possible quantum phases that may arise from particle-hole symmetry breaking. Recent advances in machine learning, particularly unsupervised learning techniques like Principal Component Analysis (PCA), offer a powerful new approach for studying complex many-body quantum systems, including FQH systems. PCA is a statistical method used to reduce the dimensionality of large datasets while preserving their most important features. This makes PCA particularly well-suited for analyzing the complex, high-dimensional wave functions that arise in fractional quantum Hall systems. Unlike traditional methods, PCA does not require prior knowledge of specific model wave functions, allowing for a more unbiased exploration of the system's phase space.

In this study, we apply PCA to investigate the effects of particle-hole symmetry breaking in the 5/2 FQH system. By introducing a model Hamiltonian that includes a three-body interaction, we explore how the ground state of the system evolves between the Pf and APf states as the strength and direction of the three-body potential are varied. PCA enables us to analyze the many-body wave functions generated by this model without relying on predefined Pf or APf states, providing a more general framework for understanding the phase transitions in the system. The results of our study not only provide new insights into the role of particle-hole symmetry breaking in the 5/2 system but also demonstrate the utility of machine learning methods for exploring complex quantum systems.

This paper is organized as follows: In Section 2, we introduce the model Hamiltonian and describe the numerical methods used to generate wave functions for PCA analysis. Section 3 details the application of PCA to the system and presents the key results, including the identification of phase transitions between topological states. Finally, in Section 4, we summarize our findings and discuss the broader implications of our work for understanding fractional quantum Hall systems and the application of machine learning techniques in condensed matter physics.

## 2. Model and Method

We consider a Single-layer two-dimensional electron gas system subjected to a perpendicular magnetic field B, the system is placed in the half-filled Landau level N=1LL, with periodic boundary conditions applied to the magnetic translation operator within a rectangular torus geometry of size $L_x * L_y$. To investigate the effect of particle-hole symmetry breaking, the three-body effect is introduced in this work. Previous studies have shown that the effective three-body pseudopotential caused by the mixing of Landau levels may be negative. Therefore, the parameter gamma controls both the strength and direction of this interaction, where the sign of gamma determines whether the interaction is attractive or repulsive. It is evident that when gamma is positive or negative, it reflects the particle-hole conjugate relationship. Specifically, if one end of gamma represents a particular state, changing the sign of gamma results in the corresponding Hamiltonian's ground state being the particle-hole conjugate of the original state. When gamma is very large and positive. As gamma changes from a large negative value through zero to a large positive value, the system's ground state transitions from the APf state to the Pf state. Since the APf and Pf states have completely different topological properties, a topological phase transition is expected to occur at gamma equals zero. The low-energy state and energy spectrum corresponding to the above Hamiltonian can be solved by exact diagonalization (ED) calculation. The emergence of these states indicates a phase transition driven by particle-hole symmetry breaking.

PCA offers a powerful approach to reduce the complexity of multidimensional data while preserving its most critical features. In the context of our FQH study, PCA is employed to analyze many-body wave functions derived from the ED of the Hamiltonian. The aim is to simplify the representation of these wave functions, allowing us to identify key components and understand how the system's ground state evolves as the strength of the three-body interaction parameter gamma changes. PCA simplifies the analysis of the wave function space by identifying the most relevant features governing the system's behavior. It highlights the symmetry-breaking effects induced by the three-body interaction and helps pinpoint critical transitions between distinct topological phases. This dimensionality reduction provides a clearer view of the underlying physics, making it an essential tool for understanding complex quantum systems.

3. **Results and discussions**

This study investigates the 5/2 FQH system by analyzing the phase transitions driven by a three-body potential that breaks particle-hole symmetry. Systems with varying numbers of filled electrons, ranging from $N_e = 8$ to $N_e = 12$, are studied, and the behavior of the wave functions is explored using PCA. The three-body potential parameter gamma ranges from -1 and 1, with wave functions sampled at increments of $\Delta\gamma=0.01$. Taking $N_e = 10$ as an example, with an aspect ratio of $L_x/L_y=0.94$, Figure 1 shows the energy spectrum obtained from ED as a function of gamma. From the figure, it can be observed that at gamma = 0, the energy spectrum exhibits symmetry, indicating that the system has particle-hole symmetry at this point. As gamma deviates from zero, the system gradually transitions toward a triply degenerate state. However, based solely on the energy spectrum, it is not possible to directly determine whether the ground state is a Pf state or an APf state.

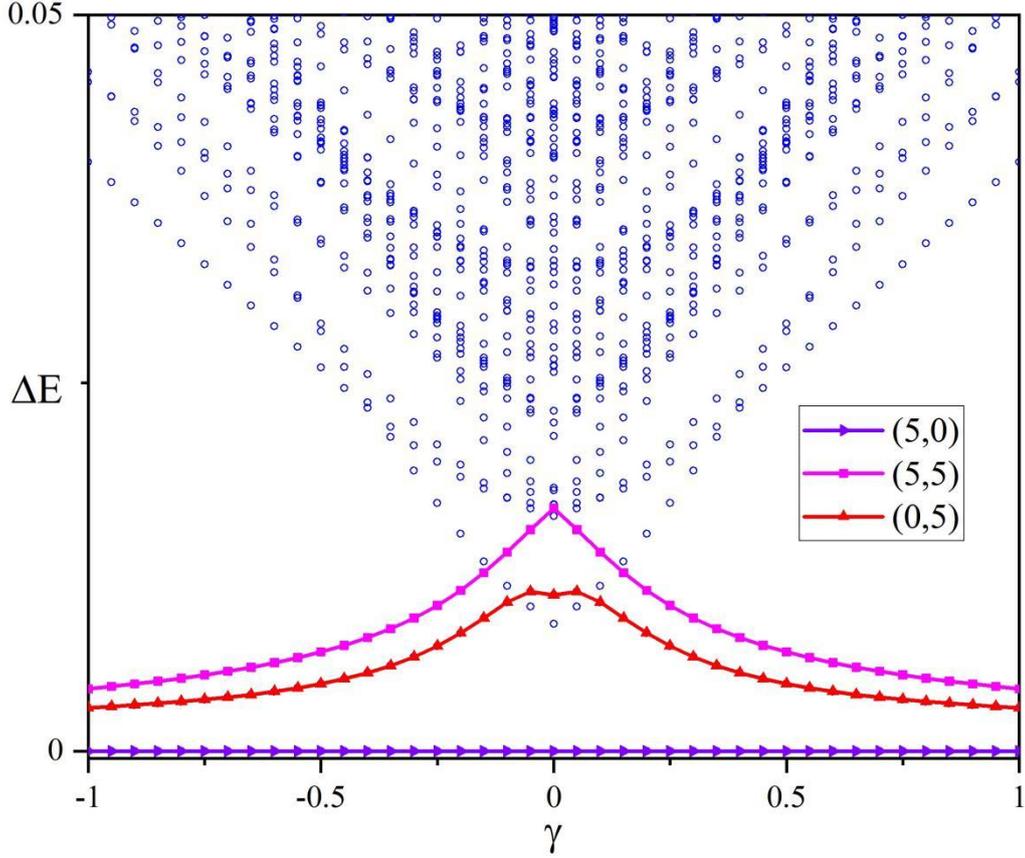

**Figure 1.** The low-lying energies as a function of gamma with $N_e = 10$ system. The purple oblique triangle, pink square and red up triangle symbol lines represent the ground states at the pseudomomenta (5,0), (5,5) and (0,5), respectively.

Taking $N_e = 10$ as an example, with an aspect ratio of $L_x/L_y=0.94$, Figure 1 shows the energy spectrum obtained from ED as a function of gamma. From the figure, it can be observed that at gamma = 0, the energy spectrum exhibits symmetry, indicating that the system has particle-hole symmetry at this point. As gamma deviates from zero, the system gradually transitions toward a triply degenerate state. However, based solely on the energy spectrum, it is not possible to directly determine whether the ground state is a Pf state or an APf state. To further investigate the properties of these ground states, we performed PCA. By analyzing the principal components of the system's wavefunction, it is possible to extract critical physical information from a low-dimensional subspace, thereby providing a clearer characterization of the ground states. This method helps to understand the evolution mechanism of the ground states and the phase transition characteristics for $\gamma > 0$ and $\gamma < 0$. Additionally, PCA can verify whether these states possess the essential features of Pf or APf states, offering a more comprehensive and detailed understanding of the system's phase transition behavior.

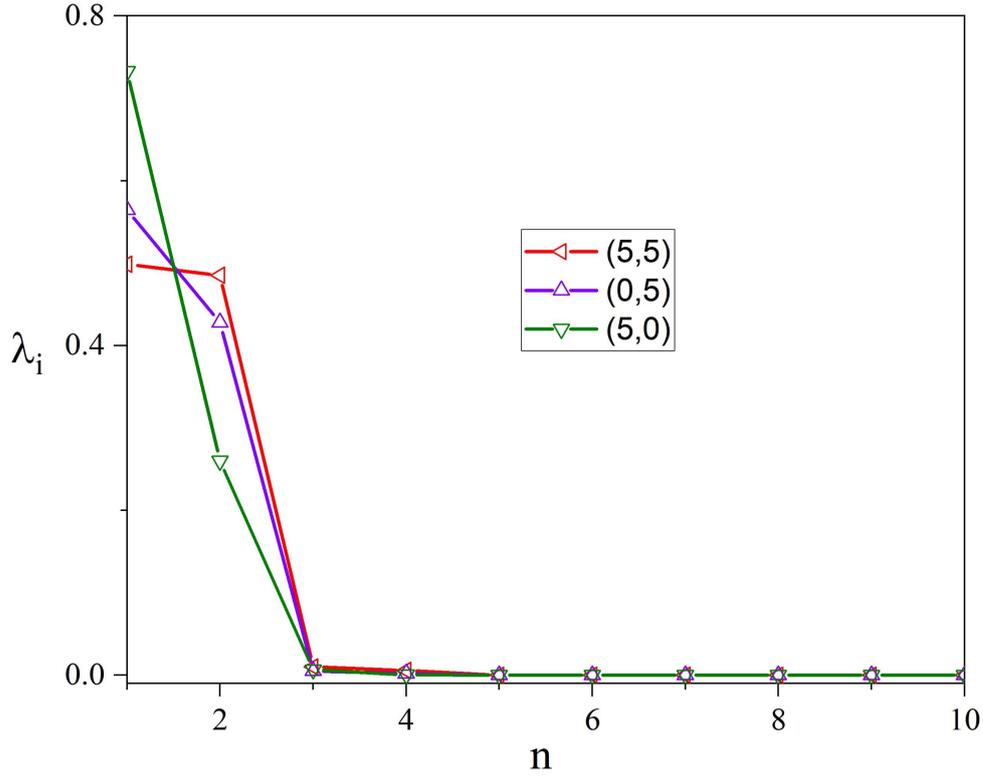

**Figure 2.** The proportion of the first ten principal components at pseudomomentum (5,0), (5,5) and (0,5) for the aspect ratio $L_x/L_y = 0.94$ with electron number $N_e = 10$. The red hollow left triangle dot line is (5, 5), the purple hollow up triangle dot line is (0, 5) and the green hollow down triangle dot line is (5, 0).

In Figure 2, we analyze the contributions of the top ten principal components at pseudomomenta (5,0), (5,5) and (0,5). The results indicate that the first and second principal components dominate the ground state wave functions. The sum of these two principal components is approximately 0.99, allowing higher components to be neglected. The dominance of the first two components demonstrates that, despite the Hilbert space has a high dimensionality, the physical description of the phase transitions can be effectively reduced to a lower-dimensional space. Based on the above analysis, we can construct a two-dimensional vector space using $V_1$ and $V_2$ as axes. Given a gamma, its wave function can be represented by the two-dimensional axis vectors $V_1$ and $V_2$. To further determine whether the wave function at $\gamma \neq 0$ corresponds to the Pf or APf state, we compute the projections of the wave function onto the APf and Pf states.

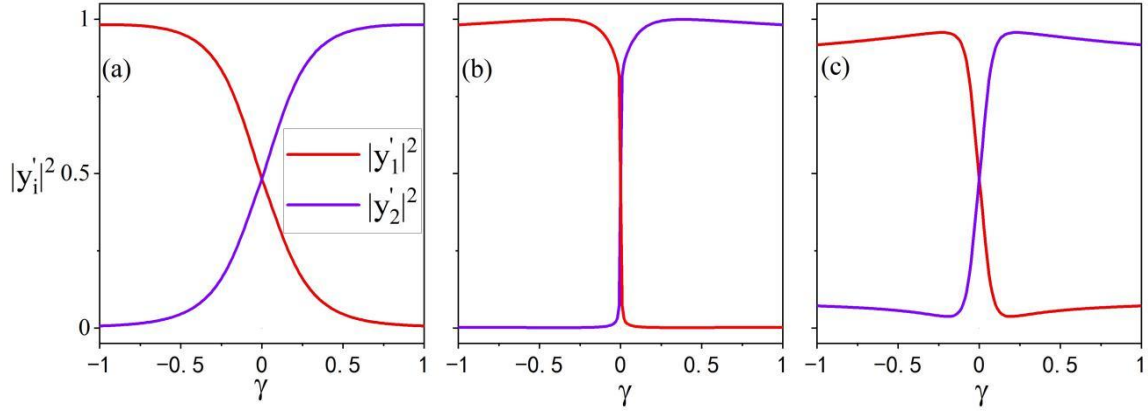

**Figure 3.** The weight of the wavefunction on APf state and Pf state varies with gamma for $N_e = 10$ system at pseudomomenta (a) (5, 0), (b) (5, 5), (c) (0, 5), respectively. The aspect ratios of the rectangular cell is $L_x/L_y=0.94$.

Figure 3 provides crucial insights into the phase transition between the Pf and APf states, driven by the gamma. The projection weights transitions from negative to positive values, illustrating the phase transition between the Pf state and the APf state and offering deep insights into how particle-hole symmetry breaking influences the system's topological phases. By analyzing the projection weights of the wave function, we observe that when gamma<0, the system primarily projects onto the APf state. As gamma increases from negative to positive, the projection weight of the APf state gradually decreases, while that of the Pf state increases. When gamma>0, the system fully transitions into the Pf state, demonstrating how its topological nature shifts as the particle-hole symmetry is progressively broken by the three-body interaction parameter gamma. Near gamma = 0, the wave function projects equally onto both the Pf and APf states, indicating a superposition of these two states due to the preservation of particle-hole symmetry. This behavior reflects the system's symmetry at the Coulomb point. At this critical point, the system exhibits a symmetric superposition of Pf and APf states, revealing a unique topological property of the FQH system at the phase transition boundary. The smooth and continuous change in the projection weights as gamma increases indicates the gradual evolution of the system's topological phase without any abrupt changes. Furthermore, the phase transition behavior varies slightly across different pseudomomenta, such as (5,0), (5,5), and (0,5). The rate of projection weight change and the smoothness near the critical point differ, suggesting that the critical dynamics may depend on the system's momentum distribution. This highlights the important role of pseudomomentum in shaping the details of the phase transition, possibly leading to different critical behaviors under varying conditions.

The physical significance of this topological phase transition is profound, particularly for topological quantum computing. Non-Abelian anyons in the Pf and APf states are regarded as potential candidates for topological qubits. By tuning the three-body interaction parameter gamma, the system can transition between the Pf and APf states, offering a mechanism for controlling topological qubits. The ability to switch between phases by adjusting gamma provides a theoretical foundation for manipulating topological qubits, potentially enabling robust quantum information processing. Although this mechanism is theoretically promising, experimental validation remains ongoing. Nonetheless, these findings provide valuable insights into the fine structure of topological phase transitions in FQH systems, enhancing our understanding of complex many-body systems and their potential applications in future quantum computing hardware development.

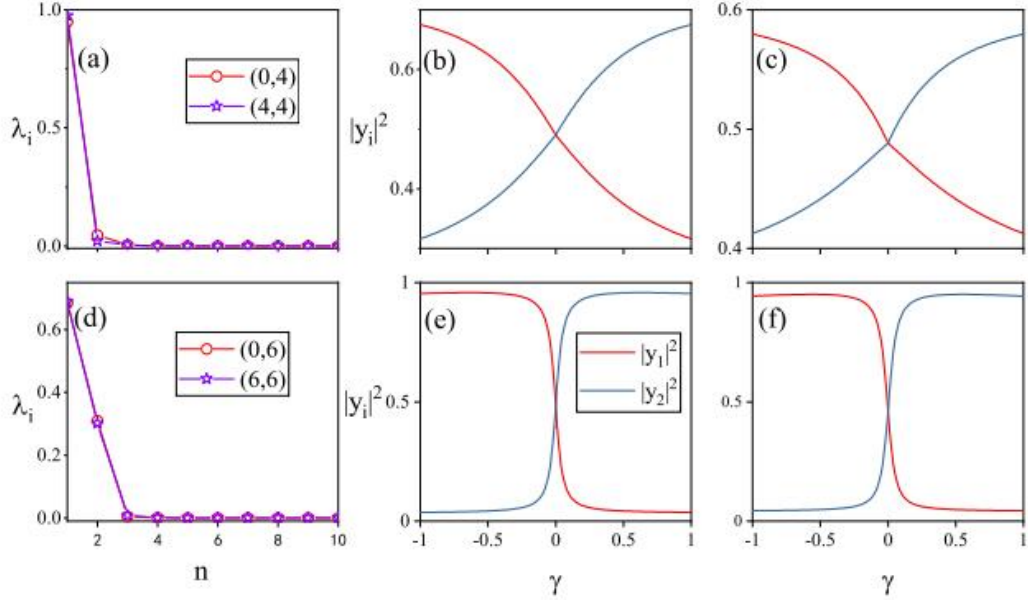

**Figure 4.** (a)The proportion of the first ten principal components at pseudomomentum (0, 4) and (4, 4) for a square cell with electron number $N_e = 8$. The red circle dot line is (0, 4) and the purple star dot line is (4, 4). The projective weights of the wave function on APf state and Pf state vary with gamma for $N_e = 8$ system at pseudomomenta (b) (0, 4) , (c) (4, 4). (d) The proportion of the first ten principal components at pseudomomentum (0, 6) and (6, 6) for $N_e = 12$. The red circle dot line is (0, 6) and the purple star dot line is (6, 6). The projective weights of the wave function on APf state and Pf state vary with gamma for $N_e = 12$ system at pseudomomenta (e) (0, 6) and (f)(6, 6).

Figure 4 presents the principal component analysis and the variation of the wavefunction projection weights with gamma for these two systems. The PCA results (Figures 4(a) and (d)) show that, for both $N_e = 8$ and $N_e = 12$ systems, the first two principal components dominate, capturing nearly all the information of the wavefunction.

This indicates that even with increasing electron numbers and the resulting significant complexity in the Hilbert space, the key physical features of the wavefunction can still be effectively captured within a low-dimensional subspace. Meanwhile, Figures 4(b), 4(c), 4(e), and 4(f) illustrate the smooth evolution of the wavefunction projection weights on the Pf and APf states with gamma. For gamma<0, the projection weights are predominantly concentrated on the APf state, while for gamma>0, the projection shifts towards the Pf state. At gamma=0, the projection weights of the Pf and APf states are equal, reflecting the central role of particle-hole symmetry at the critical point. This smooth variation confirms that the phase transition is continuous and is controlled by the three-body interaction parameter gamma.

Although the $N_e = 8$ and $N_e = 12$ systems exhibit slight differences in the transition rates and details near the critical point, their overall trends are highly consistent with the $N_e = 10$ system. This consistency further confirms that the phase transition is driven by the three-body interaction parameter gamma, and that particle-hole symmetry breaking plays a decisive role in the evolution between the Pf and APf states. These conclusions are consistent with those obtained using traditional methods. Therefore, through the analysis of the three example systems, it is demonstrated that PCA is suitable for studying phase transitions induced by particle-hole symmetry breaking.

## 4. Conclusions

In In this study, PCA is used to study the ground state variation with the three-body potential in a 5/2 fraction quantum Hall system of different sizes, which breaks the particle-hole symmetry. When the three-body potential appears, the system's ground state shifts toward either the direction of the APf state or Pf state as gamma changes. The Coulomb point of particle-hole symmetry is the symmetric superposition of APf and Pf states. At the Coulomb point, APf and Pf states each occupy equal weight, forming an equal weight superposition. This finding is qualitatively consistent with previous studies on 5/2 fraction quantum Hall systems using other three-body potential models, demonstrating the effectiveness of PCA in phase transitions induced by particle-hole symmetry breaking. It is particularly noteworthy that PCA is model-independent, offering excellent versatility and can be extended to other systems with phase transitions caused by particle-hole symmetry breaking, especially those that are not yet fully understood or lack theoretical models.

**Conflicts of interest**

There are no conflicts to declare.

**Acknowledgments**

This work was funded by the National Natural Science Foundation of China (Nos. 12347169), the Natural Science Foundation of Jiangsu Province (No. BK20240892), the China Postdoctoral Science Foundation (No. 2023M732998), the Outstanding Doctor Program of Yangzhou City Lv Yang Jin Feng (No. YZLYJFJH2022YXBS087), the numerical computations were performed on Hefei advanced computing center.